\documentclass[reprint,twocolumn,prb]{revtex4-1}

\usepackage{amsmath}
\usepackage{amssymb}
\usepackage{graphicx}
\usepackage{color}
\usepackage{bm}

\begin{document}

\title{Strong Electron-Phonon Interaction and Colossal Magnetoresistance in
  EuTiO$_3$} 
\author{Ruofan~\surname{Chen}}
\author{Ji-Chang~\surname{Ren}} 
\author{Km~\surname{Rubi}}
\author{R.~\surname{Mahendiran}} 
\author{Jian-Sheng~\surname{Wang}}
\affiliation{Department of Physics, National University of Singapore,
  Singapore 117551, Republic of Singapore}
\date{\today}

\begin{abstract}
  At low temperatures, EuTiO$_3$ system has very large resistivities
  and exhibits colossal magnetoresistance. Based on a first principle
  calculation and the dynamical mean-field theory for small polaron we
  have calculated the transport properties of EuTiO$_3$. It is found
  that due to electron-phonon interaction the conduction band may form
  a tiny polaronic subband which is close to the Fermi level. The tiny
  subband is responsible for the large resistivity. Besides, EuTiO$_3$
  is a weak antiferromagnetic material and its magnetization would
  slightly shift the subband via exchange interaction between
  conduction electrons and magnetic atoms. Since the subband is close
  to the Fermi level, a slight shift of its position gives colossal
  magnetoresistance. 
\end{abstract}
\maketitle

\section{Introduction}
\begin{figure}
  \includegraphics[scale=1]{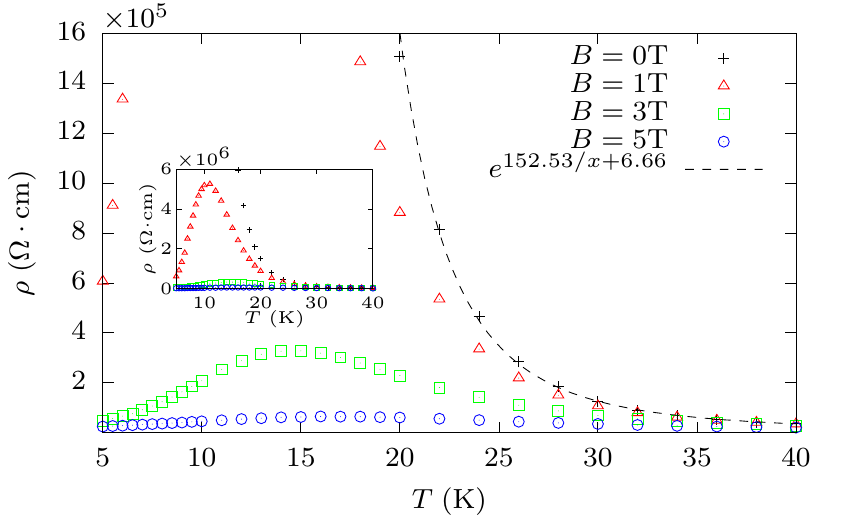}
  \caption{(color online) Experimental resistivities of EuTiO$_3$.  A
    function $y=e^{152.53/x+6.66}$ is used to fit the experimental
    resistivity without magnetic field, where
    $y=\rho/(\Omega\cdot\mathrm{cm})$ and $x=T/\mathrm{K}$.  The inset
    represents the same data but with a larger scale.}
  \label{fig:experimental-data}
\end{figure}

The colossal magnetoresistance (CMR) observed in manganites (doped
R$_{1-x}$A$_x$MnO$_3$ oxides, where R and A are trivalent rare earth
and divalent alkaline ions respectively) has attracted much attention
for the past two decades
\cite{Millis,Zang1996,Li1997,Shen1999,Zhao2000}, not only for its
possible utility in technology but also for a better theoretical
understanding of magnetoresistance. Reports on magnetoresistance in
rare earth titanates of formula RTiO$_3$ are scarce because of their
large resistivities at low temperature. Recently our experiments find
that the undoped perovskite titanium oxide EuTiO$_3$ exhibits CMR
below $\mathrm{40K}$: with the presence of an external magnetic field
its resistivity drops dramatically. In our experiments,
polycrystalline EuTiO$_3$ sample was prepared using standard solid
state reaction method in reduced atmosphere (95\% Ar and 5\%
H$_2$). More details about the sample preparation can be found in Refs
\cite{Rubi2016,Midya2016}. The DC resistivity was measured in a
Physical Property Measurement System (Quantum Design Inc, USA) using
an electrometer in two probe configuration. Experimental resistivities
are shown in Fig.~\ref{fig:experimental-data}. Due to the low
temperature and the large resistivity, whether CMR in EuTiO$_3$ is
useful in practice is not clear yet, but an investigation on it may
broaden our theoretical understanding of CMR.

When the external magnetic field is absent, the resistivity $\rho$ of
EuTiO$_3$ decreases exponentially with increasing temperature like in
a semiconductor:
\begin{equation}
  \label{eq:resistivity-semiconductor}
  \rho\approx\rho_0e^{\Delta E/k_BT},
\end{equation}
where $\Delta E$ is the gap between the bottom of conduction band and
the Fermi level, and $\rho_0$ is large and weakly temperature
dependent.  The curve of $\rho$ can be fitted by an exponential
function $e^{152.53/x+6.66}$, and substituting this fitting into
Eq.~\eqref{eq:resistivity-semiconductor} we obtain that $\Delta
E\approx 152.53\,k_B\mathrm{K}\approx0.013\,\mathrm{eV}$, which is a
small value. With such a small gap, a relatively high carrier density
is expected. However, this contradicts with large resistivities shown
in the figure, which are mostly larger than
$10^5\,\Omega\cdot\mathrm{cm}$. This paradox indicates that, rather
than applying the theory of semiconductor directly, some other factors
need to be considered.

\section{The Model}
It has been reported in literature
\cite{Millis,Zang1996,Li1997,Shen1999,Zhao2000} that
electron-phonon, especially polaronic, interaction plays an important
role in CMR. Millis \cite{Millis} pointed out that in the
La$_{1-x}$Sr$_x$MnO$_3$ double exchange alone can not explain its
resistivity, and Jahn-Teller effect must be considered. Later Zhao
\cite{Zhao2000} showed that in La$_{1-x}$Sr$_x$MnO$_3$ the electron
transport behaviour is consistent with small polaron coherent motion
which involves a strong coupling between electrons and soft optical
phonons. And the small polaron effect had also been observed in a
titanium oxide, rutile ($\mathrm{TiO_2}$), single crystal
\cite{Bogomolov}. Therefore we also take strong electron-phonon
interaction into consideration and use small polaron formalism to
model it in EuTiO$_3$ at low temperature.

The ions $\mathrm{Eu^{2+}}(4f^7)$ in EuTiO$_3$ have a large localized
spin ($S=7/2$), which is the source of magnetism. The magnetic
properties of EuTiO$_3$ can be described by an
antiferromagnetic Heisenberg model \cite{Katsufuji2001} and
Weiss mean-field theory, and the N\'eel temperature of the crystal is
about $5.4\,\mathrm{K}$ \cite{Alho2012,Ranke2012,Midya2016}. The
magnetization $\bm{M}$ is defined to be a dimensionless quantity as
$2\langle{\hat{\bm{S}}}\rangle$ and $M$ is the absolute magnitude
$|\bm{M}|$, here a Land\'e factor 2 is included. The mean-field
calculation of $M$ as a function of temperature $T$ and magnetic field
$B$ is shown as the inset of Fig.~\ref{fig:DOS0}. The
conduction electrons ($t_{2g}$ orbitals electrons of Ti) are assumed
to be coupled with magnetic atoms (Eu) via simple exchange interaction
\cite{Haas1970}, and such interaction would cause shift of the
conduction band when the system is magnetized. As will be shown later,
such shift is responsible for CMR. Note that unlike the case in the
double exchange model, the conduction electron and the magnetic atom
are not at the same site, i.e., the exchange interaction is not
intraatomic. Therefore the strength of exchange interaction is suppose
to be small and the scattering by magnetic atoms may be neglected. The
smallness of exchange interaction also implies that the material
  has a fine electronic structure if the CMR effect is due to this
interaction.

It can be also seen from the insert of Fig.~\ref{fig:DOS0}
that in the presence of magnetic field the magnetization increases,
while the resistivity drops dramatically, with decreasing temperature
and increasing magnetic field. This reminds us that the drop of
resistivities may be related to the increase of magnetization.

In this article the electron-phonon interaction is assumed to be the
Holstein model type \cite{Ciuchi1997,Fratini2003,Holstein1959}, which
means that conduction electrons are coupled with local dispersionless
optical phonons. The Hamiltonian is written in two parts as
\begin{equation}
  \hat{H}=\hat{H}_0+\hat{H}_1,
\end{equation}
where 
\begin{equation}
  \begin{array}{l}
    \displaystyle\hat{H}_0=-\sum_{ij,\alpha}t_{ij}
    \hat{c}_{i\alpha}^\dagger\hat{c}_{j\alpha}
    +\omega_0\sum_{i}\hat{a}_i^\dagger\hat{a}_i\\
    \displaystyle\phantom{\hat{H}_0=\sum}
    +g\sum_{i\alpha}\hat{c}_{i\alpha}^\dagger\hat{c}_{i\alpha}(\hat{a}_i+\hat{a}_i^\dagger)
  \end{array}
\end{equation}
is the Holstein model Hamiltonian. Here $t_{ij}$ is the hopping
  matrix element connecting site $i$ and $j$, and the operator
$\hat{c}_{i\alpha}^\dagger$ ($\hat{c}_{i\alpha}$) creates (destroys)
an electron with spin $\alpha$ at site $i$ while $\hat{a}_i^\dagger$
($\hat{a}_{i}$) creates (destroys) a phonon at site $i$. The frequency
of the optical phonon is denoted by $\omega_0$ and the coupling
strength of electron-phonon interaction is denoted by $g$. The
coupling between electrons and magnetic atoms is expressed by the
exchange interaction
\begin{equation}
  \hat{H}_1=J\sum_{i}\hat{\bm{s}}_i\cdot\bm{M}(T,B),
\end{equation}
where $J$ is the corresponding coupling strength, the operator
$\hat{\bm{s}}_i$ is electron spin operator
$\sum_{\alpha\beta}\hat{c}_{i\alpha}^\dagger\frac{1}{2}\bm{\sigma}_{\alpha\beta}\hat{c}_{i\beta}$
at site $i$ with $\bm{\sigma}_{\alpha\beta}$ the Pauli matrices vector
and $\bm{M}$ is the magnetization of the material which is a function
of temperature and magnetic field.

Back to Eq.~\eqref{eq:resistivity-semiconductor}, it should be
emphasized that although the thermal activated hopping process of
small polaron gives the same form of resistivity
\cite{Fratini2003,Mahan2000}, Eq.~\eqref{eq:resistivity-semiconductor}
is unlikely due to this process. This can be argued as follows.  There
exists a crossover temperature where the polaron motion crosses over
from band like to thermal activated hopping motion, and the hopping
process begins to dominate when temperature is above such crossover
temperature. This crossover temperature should be around $0.4\omega_0$
\cite{Ciuchi1997,Fratini2003,Mahan2000}. According to the first
principle calculation the highest frequency of optical phonon is about
$0.1\,\mathrm{eV}$ \cite{Rushchanskii2012}, and we assume it to be the
value of $\omega_0$. This value means that the crossover temperature
should be around $464\,\mathrm{K}$, which is far above
$40\,\mathrm{K}$. Besides, experiments showed that the crossover
temperature of rutile ($\mathrm{TiO_2}$) is about $300\,\mathrm{K}$
\cite{Bogomolov}, which is also far above $40\,\mathrm{K}$.

\section{The Method}

\begin{figure}
  \includegraphics[scale=1]{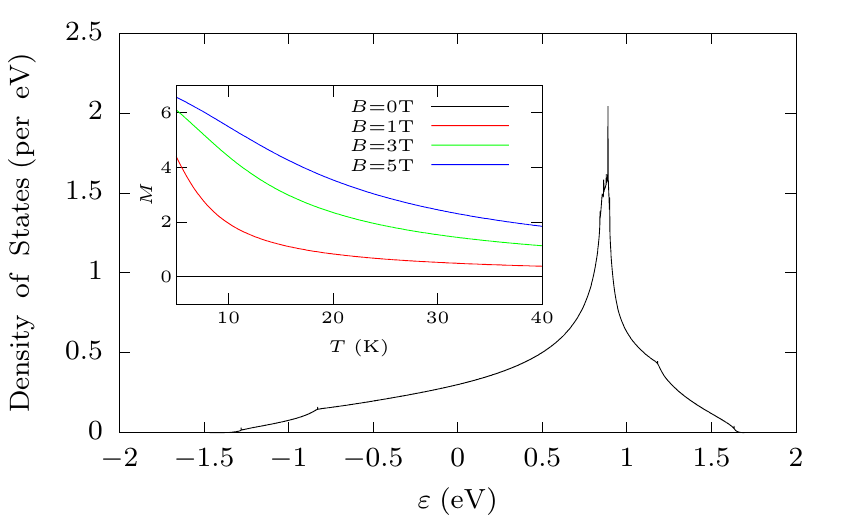}
  \caption{(color online) The DOS by first principle calculation. The
    Fermi level, which need to be fitted by experimental data later,
    is not specified here. The first principle DFT calculation is
    carried out within the spin-polarized generalized gradient
    approximation (GGA) \cite{GGA} using norm-conserving
    pseudopotentials. We use a kinetic energy cutoff of
    $60\,\mathrm{Ry}$ and a $10\times10\times10$ $\Gamma$-centered
    $k$-point mesh for the unit cell simulations. Then the mesh is
    interpolated up to $40\times40\times40$ by Wannier functions. The
    inset represents the mean-field calculation of magnetization of
    EuTiO$_3$, here $M=|\bm{M}|$ is defined as a dimensionless
    quantity.}
  \label{fig:DOS0}
\end{figure}

To obtain the electronic structure of a specified material which
involves strong interaction, the method combining first principle
calculation and dynamical mean-field theory (DMFT) are often used
\cite{Georges1996,Kotliar2006}. In this article the density of states
(DOS) of conduction band ($t_{2g}$ orbitals of Ti atom) is obtained
via first principle calculation. Then based on this DOS we apply the
DMFT for the small polaron \cite{Ciuchi1997,Fratini2003}, which is a
method for Holstein model, to handle electron-phonon interaction.
According to the DMFT results, due to electron-phonon coupling it is
possible to form a tiny polaronic subband of the conduction band near
the Fermi level. Such a tiny subband would reduce both the carrier
density and electrical mobility. The existence of such a tiny subband
may be the explanation of the coexistence of small $\Delta E$, low
carrier density and high resistivity.

\begin{figure}
  \includegraphics[scale=1]{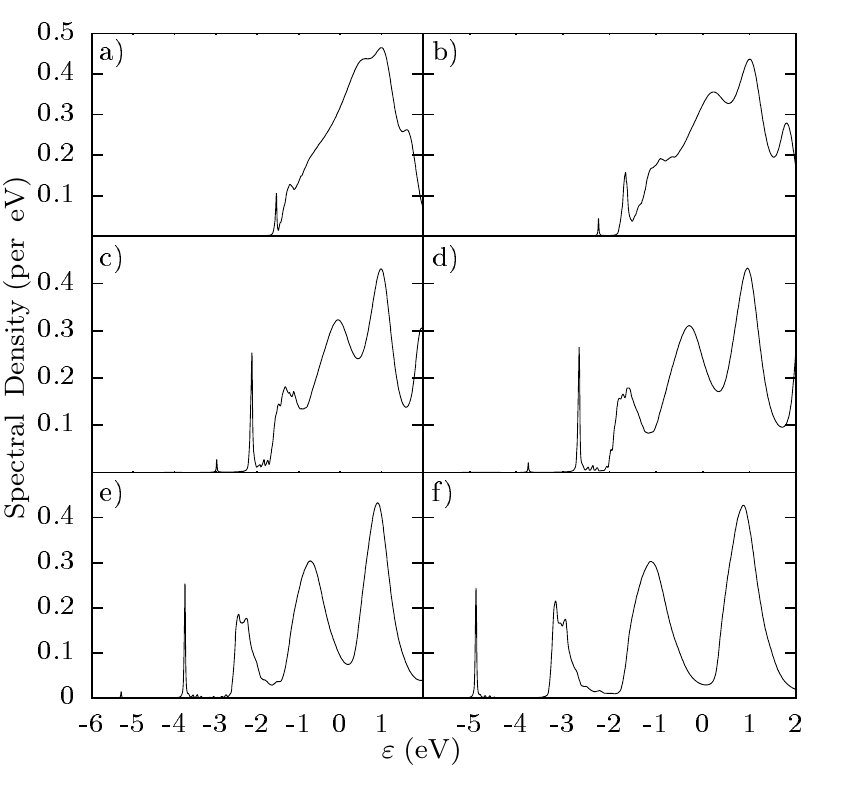}
  \caption{The spectral density calculated by DMFT with $g=$ (a)
    0.6\,eV, (b) 0.8\,eV, (c) 1.0\,eV, (d) 1.2\,eV, (e) 1.6\,eV, and
    (f) 2.0\,eV. The unperturbed DOS used by DMFT is shown in
    Fig.~\ref{fig:DOS0}.}
  \label{fig:final-dos}
\end{figure}

The carrier density is assumed to be very small and the temperature
under consideration is low, therefore we can use the zero temperature
formalism of DMFT for small polaron which works in the extreme dilute
limit \cite{Ciuchi1997} to deal with $\hat{H}_0$.  In this formalism
an analytic impurity solver can be built by continued-fraction
expansion \cite{Viswanath1994}. The crucial advantage of such an
impurity solver is that it allows the DMFT procedure to be done
directly in real frequency domain and thus no analytic continuation is
needed.  The unperturbed DOS before the DMFT procedure is calculated
via density functional theory (DFT) by Quantum Expresso \cite{QE} and
then interpolated by Wannier functions
\cite{Marzari1997,Souza2001}. The DOS of $t_{2g}$ orbitals of Ti atom
thus calculated is shown in Fig.~\ref{fig:DOS0}. After the DMFT
calculation for $\hat{H}_0$, an energy dependent self-energy
$\Sigma_0(\varepsilon)$ and the corresponding retarded Green's
function $G_0(\varepsilon)$ are obtained. As mentioned earlier, the
frequency of the optical phonon $\omega_0$ is about
$0.1\,\mathrm{eV}$. The spectral density calculated by DMFT with
$\hat{H}_0$ for different values of $g$ is shown in
Fig.~\ref{fig:final-dos}, where the spectral density is given by
$-\frac{1}{\pi}\mathrm{Im}\,G_0(\varepsilon)$.

It can be seen from Fig.~\ref{fig:final-dos}(a) that when $g$
increases to $0.6\,\mathrm{eV}$ a small peak appears at the bottom of
the band.  As $g$ goes to $0.8\,\mathrm{eV}$ a second peak appears and
the first one becomes lower, see Fig.~\ref{fig:final-dos}(b). It can
be also seen from (c)--(e) that when $g$ becomes larger, the first
peak remains but its position is shifted.  And in (f) the first peak
is shifted outside the figure.

The first two peaks can be treated as two tiny subbands of the
conduction band, and they can provide conduction electrons.  At first
glance, the first subband is too small and may be neglected. However,
our calculation of resistivities shows that the second subband still
provides too many conduction electrons for such large
resistivities. Thus we focus on the first subband which is the
  polaronic subband. If this subband is close to the Fermi level,
then it can explain the smallness of $\Delta E$. And since the subband
is tiny, the carrier density would be still low.


Now let us turn to $\hat{H}_1$. The magnetization $\bm{M}$ in
$\hat{H}_1$ is an average quantity, which means that scatterings due
to localized spins are neglected.  It is easy to see that $\hat{H}_1$
would not change the shape of electronic band structure, but only
shift the self-energy according to different spins of
electrons. Therefore the final results for self-energy is
$\Sigma_\alpha=\Sigma_0\pm\frac{1}{2}JM(T,B)$ with $M=|\bm{M}|$ for
spin up and down respectively. This is a kind of band shift.  The
final Green's function $G_\alpha$ would change according to the band
shift for different spin $\alpha$. And the final spectral density is
then given by $-\frac{1}{\pi}\mathrm{Im}\,G_\alpha$.

The static conductivity, which is the inverse of resistivity, can be
calculated via the Kubo-Greenwood
\cite{Greenwood1958,Kubo1957,Economou2006} formula (here the trace
contains spin summation)
\begin{equation}
  \sigma=\frac{e^2\hbar}{\pi V}\int\left(-\frac{\partial f}{\partial\varepsilon}\right)
  \mathrm{Tr}[\hat{v}_x\mathrm{Im}G(\varepsilon)\hat{v}_x\mathrm{Im}G(\varepsilon)]
  d\varepsilon,
\end{equation}
where $V$ is the volume of system and $\hat{v}_x$ is the operator for
a component of velocity. Here we can use Boltzmann distribution
$f=\exp[(\mu-\varepsilon)/k_BT]$ instead of the Fermi-Dirac
distribution since the temperature is much lower than $\Delta E$. Due
to the band shift the distribution function becomes
$f=\exp[(\mu-\varepsilon\mp\frac{1}{2}JM)/k_BT]$. The band with spin
down is shifted by $-\frac{1}{2}JM$, thus it goes closer to the Fermi
level and provides more conduction electrons. While another band with
spin up would be shifted away from the Fermi level and the carrier
density in it would be reduced. However, because the distribution
function is exponential, the total carrier density increases and the
resistivity decreases accordingly. An important point here is that
$\Delta E\approx0.013\,\mathrm{eV}$ is very small, thus even a small
amount of shift, say $30\,k_B\mathrm{K}\approx0.0026\,\mathrm{eV}$,
would cause an obvious difference. While in other materials such a
small amount of shift may be just ignored. This is the origin of CMR.

\section{Results and Discussion}
\begin{figure}
  \includegraphics[scale=1]{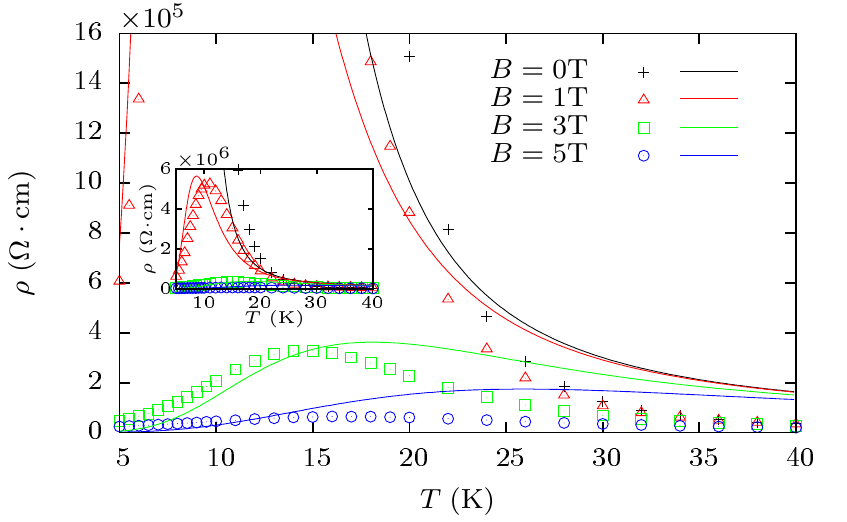}
  \caption{(color online) Resistivities of EuTiO$_3$. Solid lines
    represent theoretical results, and experimental data (dots) are
    plotted here for comparison. The inset represents the same data
    but with a larger scale.}
  \label{fig:calculated-resistivity}
\end{figure}

We calculate resistivities based on the first peak in
Fig.~\ref{fig:final-dos}(c) with $g=1.0\,\mathrm{eV}$. This value of
$g$, of course, may not be accurate for real situation, so we need to
adjust our parameters to fit experimental data. More discussion about
the value of $g$ can be found in appendix. We set the Fermi level at
$-3.0778\,\mathrm{eV}$. Note that because the carrier density is very
sensitive to the band shift, the position of the Fermi level need to
be carefully placed. With this Fermi level the electron occupation
number per site at $20\,\mathrm{K}$ without magnetic field is about
$8.457\times10^{-7}$, which is consistent with the extreme dilute
limit used by DMFT. The group velocity $v_x(\bm{k})$ of electron is
obtained by our first principle calculation. The maximum velocity is
about $10^5\,\mathrm{m/s}$. The lattice constant of EuTiO$_3$ is about
$4\times10^{-8}\,\mathrm{cm}$ \cite{Katsufuji2001}. And the value of
$J$ is set equal to
$0.0025\,\mathrm{eV}\approx29\,k_B\mathrm{K}$. Resistivities
calculated by Kubo-Greenwood formula are shown as solid lines in
Fig.~\ref{fig:calculated-resistivity}.

It can be seen that this value of $J$ fitted by experimental data is
indeed small, this confirms our assumption. And because the tiny
subband is quite close to the Fermi level so such a small $J$ still
has a strong effect on the resistivity. Shapes of curves in
Fig.~\ref{fig:calculated-resistivity} are basically controlled by the
distribution function with band shift
$f=\exp\{[\mu-\varepsilon\mp\frac{1}{2}JM(T,B)]/k_BT\}$. In fact, due
to the band shift, the carrier density can be written in a form
$n(T,B)=n_0(T)e^{\mp\frac{1}{2}JM/k_BT}$, where $n_0(T)$ is the
carrier density without magnetic field. With this form of $n$,
experimental data can be simply fitted by Einstein formula
$\sigma=neb$ with a constant mobility $b$. The problem of this simple
fitting is that either $n_0$ or $b$ need to be very small. Now the
existence of the tiny polaronic subband reduces both the carrier
density and electrical mobility. This simple fitting is also an
evidence that the band shift is the origin of CMR for EuTiO$_3$.

In summary, we have applied DFT+DMFT method to calculate the
electronic structure of $t_{2g}$ orbitals of Ti atom in EuTiO$_3$.
Based on this electronic structure we have calculated the transport
properties of EuTiO$_3$ and explained the CMR in it. It is found that
due to polaronic interaction the conduction band can form a tiny
subband. This subband may be close to the Fermi level and responsible
for conduction electrons. Since the subband is very small, the carrier
density is still very low even it is close to the Fermi level. This is
the reason why resistivities of EuTiO$_3$ are quite high. Conduction
electrons are also coupled with magnetic atoms via exchange
interaction, and this interaction would slightly shift the electronic
band when the material is magnetized. And because the subband is close
to the Fermi level, a slight shift is enough to cause CMR.

It is clear that this mechanism occurs in semiconductor and involves
no strong intraatomic exchange interaction as in the double exchange
model. Unlike in La$_{1-x}$Sr$_x$MnO$_3$ system which is metallic
\cite{Millis,Zang1996,Li1997,Shen1999,Zhao2000}, the change
of carrier density caused by band shift plays a main role in the CMR
of EuTiO$_3$. 
Besides, because at low temperature the carrier
density for different electron spin changes dramatically when material
is magnetized, EuTiO$_3$ has a potential for spintronic device.

Our model is a simplified model. It is not enough to obtain really the
fine electronic structure of EuTiO$_3$, thus the agreement with
experimental data remains at a qualitative level. A more careful
treatment on first principle calculation and DMFT procedure may
improve the accuracy. What's more, experiments which can measure the
carrier density change for different spin respectively, or just the
total density change, in the presence of magnetic field can help to
verify or falsify the validity of our theoretical description.

This work is supported by MOE tier 2 grant R-144-000-349-112.

\appendix
\section{The Value of Parameters}
Here we shall discuss some details about the values of parameters
$\omega_0$ and $g$. 

It has been mentioned earlier that the value of $\omega_0$ is assumed
to be the highest frequency optical phonon. The main reason is that
the highest phonon band is well separated with other bands and is
relatively flat. The flatness of the band indicates that the band is
relatively local, which is consistent with the assumption of Holstein
model.

The value of $g$ is chosen to be $1\,\mathrm{eV}$, it should be noted
that this value is a large value for electron-phonon
interaction. Especially, applying Lang-Firsov \cite{Lang1962}
transformation, which is the standard method for small polaron theory,
on Holstein model yields some unphysical polaron parameters. The
bandwidth renormalization constant for small polaron is
$\exp(-g^2/\omega_0^2)=\exp(-100)=3.72\times10^{-44}$, which means the
bandwidth of polaron subband would be at the order of $10^{-44}$ and
thus this subband would be so fragile that it would be immediately
washed out in a real material. However, Lang-Firsov transformation
also shows the position of small polaron subband should be located
around $-g^2/\omega_0=-10\,\mathrm{eV}$, which is far from the subband
we obtain. Therefore what we obtain is not the fragile polaron subband
but another relatively robust subband caused by strong electron-phonon
interaction.

So is this large $g$ possible? Our first principle calculation shows
it is indeed possible in EuTiO$_3$ system.

\begin{figure}[!htbp] \centering
  \includegraphics[scale=1]{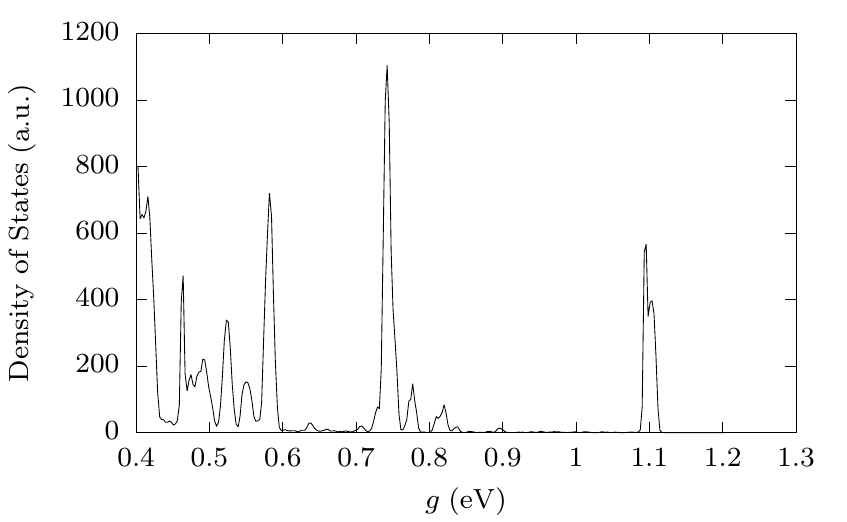}
  \caption{Density of states of the coupling constants between
electronic states with localized LO mode phonon states.}
  \label{fig:gmatrix}
\end{figure}

The DFT calculatons are performed using Quantum ESPRESSO package
\cite{QE}. The Troullier-Martins norm-conserving pseudopotentials with
the Perdew-Burke-Ernzerhof (PBE) exchange-correlation functionals
\cite{GGA} are employed to describe the interactions between valence
electrons in our system. The cutoff energies of plane waves are chosen
as $80\,\mathrm{Ry}$. A $20\times20\times20$ Monkhorst-Pack $k$-point
mesh is used for electronic self-consistent field calculations and a
$4\times4\times4$ Monkhorst-Pack $k$-point mesh is used for phonon
calculations. The convergence threshold of energy is set to be
$10^{-14}$ Ry for electron, while for phonon calculations, the
threshold is set to be $10^{-18}$ Ry to get a better convergence. The
electron phonon coupling matrix is calculated by applying formula
\cite{Giustino2017}:
${g}_{mn\nu}(\bm{k},\bm{q})=\langle u_{m\bm{k}+\bm{q}}|
\Delta_{\bm{q}\nu} v^{\rm KS}|u_{n\bm{k}} \rangle$,
where $u_{n\bm{k}}$ is the lattice-periodic function in Bloch
wavefunction, the bra and ket indicate an integral over one unit cell,
and the operator $\Delta_{\bm{q}\nu} v^{\rm KS}$ is the derivative
with a coefficient of the self-consistent potential with respect
to a collective ionic displacement corresponding to a phonon with
branch index $\nu$ and momentum $\bm{q}$. In order to get a densier
mesh to calculate the electron phonon coupling matrix, we apply
Wannier interpolation technique, as implemented in EPW code
\cite{Ponce2016}. After Fourier transformation back into momentum
space, we obtain a dense $40\times40\times40$ $\bm{k}$-point mesh for
states of electron and $40\times40\times40$ $\bm{q}$-point mesh for
states of phonon.

The DFT results can be found in Fig.~\ref{fig:gmatrix}, it can be seen
that the highest values of the elements of the electron-phonon
coupling matrix elements are around $1.1\,\mathrm{eV}$. Since Holstein
model is used, in which electrons are coupled with localized phonons,
we focus on the coupling between electrons and phonons of highest
longitudinal optical mode. Therefore in Fig.~\ref{fig:gmatrix} only
those results of LO mode with coupling constant larger than
$0.4\,\mathrm{eV}$ are represented. These results show that the value
of $g$ can reach about $1.1\,\mathrm{eV}$, and thus our value is
consistent with the DFT results.

\section{The Extreme Dilute Limit}
In a strongly correlated system, usually the value of electron
occupation number would greatly affect the electronic spectral
density. For instance, the DMFT results for La$_{1-x}$Sr$_x$MnO$_3$
system \cite{Millis} differ much around half filling situation for
different occupation numbers.

However, in our calculations the rigid band approximation is applied,
i.e., the spectral density remains unchanged when carrier density
changes. Here we shall explain the reason why we can adopt this
approximation.

It has been mentioned earlier that the electron occupation number per
site at $20\,\mathrm{K}$ without the magnetic field is about
$8.457\times10^{-7}$, and such small occupation number enables us to
apply extreme dilute limit and single electron approximation used in
DMFT for small polaron. With the presence of an external magnetic
field, the occupation number increases dramatically. However, even the
occupation increases $1000$ times, it is at an order of $10^{-4}$ which
is still very small. Therefore we can say that during CMR the
occupation number, although dramatically changes, is always small
enough to apply extreme dilute limit and single electron
approximation, and so the rigid band approximation. This is also an
important difference between EuTiO$_3$ system and
La$_{1-x}$Sr$_x$MnO$_3$.

\end{document}